\documentstyle[12pt]{article}

minus 1pt
\title{\bf Direct Calculations of the Odderon Intercept in the
Perturbative
 QCD}
\author{M.A.Braun\\ Department of High Energy Physics,
University
 of St. Petersburg,\\ 198904 St. Petersburg, Russia\\
P.Gauron and B.Nicolescu\\
LPTPE, Universit\'e Pierre et Marie Curie,\\
4 Place Jussieu, 75252 Paris Cedex 05,
France}
\date{}
\def\beq{\begin{equation}}
\def\eeq 
{\end{equation}}
\def\noi{\noindent}

\begin{document}
\maketitle
\medskip
\noi{\bf
Abstract.}
 The odderon intercept is calculated directly, from its expression via
an average energy of the odderon  Hamiltonian,
 using both  trial wave functions in the variational approach
and the wave function recently constructed by R.A.Janik and J.Wosiek.
 The results confirm their reported value for the energy.
  Variational calculations give energies 
 some 30\% higher. However they also predict the odderon intercept
to be quite close to unity. In fact, for realistic values of $\alpha_s$,
the intercept calculated variationally is at most 2\% lower than the
exact one: 0.94 instead of 0.96.
It is also found that  the solution for
$q_3=0$ does not belong
 to the odderon
 spectrum. The diffusion parameter is
 found to be of the order $0.6$.
\vspace{3.5cm}

\noi{\Large\bf LPTPE/UP6/10
\hspace{5cm}July 1998} 
\newpage
\section{Introduction}There has recently been much interest in the
odderon, both from experimental and theoretical points of view. On the
experimental side, various processes mediated by a $C$-odd object with an
intercept around unity
 are planned or already under investigation at HERA: the pseudoscalar
 meson production in ep collisions offers a direct probe for the
 odderon [1]. On the theoretical side, 
estimates of the relevant cross-sections have been made, using  simplest 
model for the odderon, just a $C$-odd state of three non-interacting gluons
 [2] 
or a non-perturbative Regge pole at $j=1$ [3]. Parallel to 
this there has been much activity in studying the gluon interaction effects, 
which presumably change the odderon intercept from exactly unity. In the
 perturbative QCD approach the odderon is a C-odd state of three
 reggeized gluons, which interact pairwise with a
 well-defined potential. 
The relevant equation is known since long ago [4]. It was shown that it is
 conformal invariant and splits into a pair of equations for the holomorphic
 and
antiholomorphic parts of the wave function in exactly the same manner as for
 the
pomeron [5]. An operator, later called $\hat{q}_3$, was also found, which,
 on the one hand, commutes with the odderon Hamiltonian $H$ and, on the other
hand,
 has a much simpler form than $H$, thus opening a way to simplify the solution
 of
 the odderon problem considerably [6]. Finally, much effort has been put into
 directly relating the operators $\hat{q}_3$ and $H$ to avoid using explicit
 wave functions
in search of the odderon intercept [7-10]. Following this latter
 line of approach
R.Janik and J.Wosiek (JW) calculated the  odderon energy for arbitrary 
complex eigenvalue of $\hat{q}_3$ in [8]. In a later publication [9] they
diagonalized the $\hat{q}_3$ operator and found both the odderon intercept and
its wave function.
 If one relates the odderon intercept
to the odderon "energy" $\epsilon$ as
 \footnote
{Different autors use different definitions of the odderon energy and therefore
of its relation to the intercept. This has to be kept in mind comparing
our numerical values with other references.}
\beq
\alpha_O(0)=
1-(9\alpha_s/2\pi)\epsilon,
\eeq
the result of JW for the ground state
is
\beq
\epsilon=0.16478...
\eeq
Thus they confirm our old conclusion that the
perturbative odderon intercept
 lies slightly below unity [11].

However a discussion about the validity of their
procedure  relating $\hat{q}_3$ and $H$ is still going on [12]. In
particular, for the eigenvalue
$q_3=0$ the result
which follows from their procedure is contested [12] (see also Sec. 5).

In view of this dispute a direct calculation of the odderon intercept, which
starts from its explicit expression in terms of the wave function, aquires
certain importance. It resolves in a unique manner any ambiguities involved
 in formal relations between ill-defined operators 
$\hat{q}_3$ and $H$
and associated with boundary conditions to be imposed on their respective
 eigenfunctions. Reporting on these direct calculations in this paper, we also
present two independent   variational results, which can be obtained starting
 from the
 direct expression of the odderon energy in terms of its wave function. They
 present some interest, since the functions determined in a variational
 procedure may be a good approximation to the exact one, which is quite
 complicated and ill-suited for practical calculations.

Our direct calculations confirm the value (2) found by JW to be the ground
 state odderon energy. Our  variational calculations 
with two different forms of the trial function give 
\beq
\epsilon=0.223\ {\rm and}\  \epsilon=0.226
\eeq
that is, some 30\% higher value for the energy.
This change in $\epsilon$ corresponds to a 2\% change in the odderon
 intercept, for realistic values of $\alpha_s$.
 From our calculations it also follows that the  value $q_3=0$ does not lead
 to any physical odderon state.

The paper is organized as follows. In Sections 2 and 3 we review the basic
formulas which serve as a starting point for our calculations. Sections 4 and
 5 present the results of direct calulations of the odderon energy with a
 variational and exact wave functions respectively. Some conclusions are 
drawn in Section 6.
\section{The odderon energy}
 The odderon energy can be sought as a ratio
\beq
\epsilon=E/D
\eeq
where  $E$
 and $D$ are energy and normalization functionals, quadratic
in the odderon wave
function $Z(r,\phi)$ [13].
Explicitly
\beq
E=\sum_{n=-\infty}^{\infty}\int_{-\infty}^{\infty}d\nu\epsilon_{n}(\nu)
|\alpha_{n}(\nu)|^{2}
\eeq Here
\beq
\epsilon_{n}(\nu)=2\
{\mbox
Re}\,\left(\psi(\frac{1+|n|}{2}+i\nu)
-\psi(1)\right),
\eeq
$\alpha_n(\nu)$ is 
a double Fourier transform of
$Z(r,\phi)$:
\[
\alpha_{n}(\nu)=\int_{0}^{\infty}dr
r^{-2-2i\nu}\int_{0}^{2\pi} d\phi
e^{-in\phi}\]\beq
\left(i\nu+\frac{n+1}{2}+re^{i\phi}(h-i\nu-\frac{n-1}{2})\right)
(i\nu-\frac{n-1}{2})(-\tilde{h}+i\nu-\frac{n-1}{2})Z(r,\phi)
\eeq
where
$
h $ and
$ \tilde{h} $ are the two conformal weights for the holomorphic and
antiholomorpic parts of the wave function. They have a general form
\beq
h=1/2+n-i\nu,\ \tilde{h}=1/2-n+i\nu,\ -\infty<\nu<\infty,\ \ n=\ldots -1,0,1,\ldots
\eeq
For the lowest branch of the odderon spectrum, which leads to the highest
intercept, supposedly $n=0$. This will always be assumed in the following.
For the ground state also $\nu=0$.
$D$ is obtained substituting  $\epsilon_n(\nu)$ by unity in (5).

The odderon wave function $Z$ can be considered as a function of $z=r\exp
(i\phi)$ and its conjugate. It has a
form
\beq
Z(z,z^*)=|z(1-z)|^{2h/3}\Phi(z,z^*)
\eeq where function $\Phi$ has to
be invariant under substitutions
\beq z\rightarrow 1-z,\ \ z\rightarrow 1/z,
\eeq
which follows from the requirement of Bose symmetry in the three gluons.

Eqs. (4), (5) present a direct way of calculating the odderon energy once its
 wave function is known. It also opens up a possibility of finding $\epsilon$
 by variational techniques minimizing the value of (4) in a given space of
 functions which should satisfy (9) and (10). Note that if one drops the
 restricting condition of Bose symmetry (10), then function $Z$ and 
consequently $\alpha_n(\nu)$ become abritrary, with the only requirement
 that (5) exist. 
Then it becomes clear that the minimal value of $E/D$ is realized by
$|\alpha_n(\nu)|^2=\delta_{n0}\delta(\nu)$, which gives
$\epsilon=\epsilon_0(0)=-4\ln2$, that is the pomeron energy. Since the 
space of functions obeying (10) is smaller, one gets an evident bound [14]
\beq
\epsilon>-4\ln2
\eeq
In fact this bound is very crude (cf. (2)).

\section{The $\hat{q}_3$ operator}
An operator $\hat{q}_3$ which
commutes with the odderon Hamiltonian was found in [6]:
\beq
\hat{q}_3^2=-r_{12}^2r_{23}^2r_{31}^2q_1^2q_2^2q_3^2 ,   \ \
[\hat{q}^2_3,H]=0.
\eeq
Evidently, the odderon ground state (nondegenerate) should also be an
eigenstate for $\hat{q}^2_3$. But, in contrast to the odderon Hamiltonian
 $H$, the operator
$\hat{q}_3^2$
is a finite order differential operator, which does not contain logarithms of
neither momenta nor coordinates. It  splits  into a product of two
differential operators $\hat{q}_3$ and its conjugate, each of the third order in 
complex variables $z$ and $z^*$ respectively. In this section we closely
follow the derivation of the spectrum of $\hat{q}_3$ given in [9].
 The eigenvalue equation for $\hat{q}_3$ and its conjugate
can easily be obtained using the explicit form (12). 
For the holomorphic part
 of the function $\Phi$ (see Eq. (9)) it reads [6,9] 
\begin{equation}
a(z) {d^3 \over d z^3}\Phi(z) + b(z){d^2\over d z^2} \Phi(z)
+c(z){d\over dz}\Phi(z) + d(z) \Phi(z)=0, 
\end{equation}
 where
\begin{eqnarray}
a(z)=z^3(1-z)^3,\;
 b(z)=2z^2(1-z)^2(1-2z),&&\nonumber\\
c(z)= z(z-1)\left(
z(z-1)(3\mu+2)(\mu-1)+3\mu^2-\mu\right)
,&&\nonumber \\ d(z)=
\mu^2(1-\mu)(z+1)(z-2)(2z-1) -i q_3 z (1-z)&&
\nonumber\end{eqnarray}
and $\mu=h/3$.
 
This
is a third order linear differential equation 
with  three regular singular
points at $z=0,1$ and
$\infty$.
 It has three linearly independent solutions
$u^{(0)}_i$, $i=1,2,3$
which can be chosen so as to possess a given behaviour in the vicinity
of $z=0$:
\beq
u^{(0)}_1(z)\sim z^{2h/3},\ \ u^{(0)}_2(z)\sim z^{1-h/3},\ \ 
u^{(0)}_3\sim z^{1-h/3}\log z+az^{-h/3},\ \ z\rightarrow 0
\eeq
obtained from the characteristic equation corresponding to
(13) at small $z$.

The final eigenfunction $\Phi(z,z^*)$ should be constructed as a sum of
 products of
these solutions with those for the antiholomorhic part.
\beq
\Phi(z,z^*)=\sum_{i,k=1}^{3}\bar{u}_i^{(0)}(z^*)A_{ik}^{(0)}u^{(0)}_k(z)\equiv
\bar{u}^{(0)}A^{(0)}u^{(0)}
\eeq
 Apart from the Bose
symmetry requirements (10) it has also to be a single-valued function
in the complex
$z,z^*$ plane. This latter condition puts evident restrictions on the form
of the coupling matrix $A^{(0)}$. From the behaviour (14)
one finds that the function $\Phi$ will be uniquely defined in the vicinity
of $z=0$ if and only if
\beq
A^{(0)}_{12}=A^{(0)}_{21}=A^{(0)}_{13}=A^{(0)}_{31}=0,\ \
A^{(0)}_{23}=A^{(0)}_{32}
\eeq

To have uniqueness around the two other singular points, 1 and $\infty$,
one has to know the behaviour of (15) in their vicinity. Due to the symmetry
of the Eq. (13) under the substitutions
\beq
z\rightarrow 1-1/z,\ \ z\rightarrow 1/(1-z)
\eeq
one can easily construct two other systems of solutions $u_i^{(1)}$ 
and $u^{(\infty)}_i$, $i=1,2,3$ with the same behaviour around $z=1$
and $z=\infty$ respectively. They can be expressed as superpositions of
the initial solutions,say,
\beq u^{(1)}_i(z)=\sum_{k=1}^{3}R^{(10)}_{ik}u^{(0)}_k(z)\eeq
and similarly for $u^{(\infty)}(z)$. The "transfer matrix" $R^{(10)}$ from
the solutions
$u^{(0)}$ to solutions $u^{(1)}$, 
 is
a constant matrix, which can technically be calculated once the solutions
$u^{(1)}$ and $u^{(0)}$ are known. Evidently, in terms of $u^{(1)}$
the wave function $\Phi$ can be written as
\beq
\Phi(z,z^*)=\bar{u}^{(1)}A^{(1)}u^{(1)}
\eeq
where
\beq
A^{(1)}=(R^{(01)})^T\,A^{(0)}\,R^{(01)}, \ \ R^{(01)}=(R^{(10)})^{-1}
\eeq
 For $\Phi$ to be a single valued
function of $\phi$ in the vicinity of $z=1$ it is necessary that $A^{(1)}$ 
has the same properties (16) as the matrix $A^{(0)}$. Moreover the Bose
symmetry requires that these matrices coincide. This gives an equation 
\beq
A= (R^{(01)})^T\,A\,R^{(01)}
\eeq for a matrix $A$ satisfying (16) [9]. It determines both the eigenvalues
$q_3$ and non-zero elements of the matrix $A$, that is, the eigenfunction
$\Phi(z,z^*)$ according to (15).
 Note that there is no need to
additionally 
require that $\Phi$ should be a single valued function of $\phi$
around $z=\infty$, since a contour encircling this point can be made of two 
contours around $z=0$ and $z=1$.

In [9] the transfer matrix $R$ was determined numerically from the
solutions $u$ calculated as a series in powers of $z$. Solution of (21) 
then determined the eigenvalues of $\hat{q}_3$. They are all pure imaginary.
The one corresponding to the odderon ground state turned out to be 
\beq
q_3=-0.20526\,i
\eeq
After that the value (2) for the odderon energy was obtained by JW
using their procedure  to relate the eigenvalues of $\hat{q}_3$ and $H$
constructed in [8].

Our  calculations use the form of the wave function (15) with the
matrix $A$ determined by Eq. (21) as an input to be substituted into
(5), which gives the value of the odderon energy directly.

\section{Variational calculations}
\subsection{Starting point
[13,14]}
 The first calculations of the odderon energy starting from Eqs. (4), (5)
 were done in the variational technique. In fact, taking some trial function 
which satisfies (10) and putting it into (4) and (5) one obtains an upper
 bound for the odderon energy and thus a lower bound for its intercept. 
Of course the problem consists in choosing a good trial function, on the one
 hand, close enough to the
exact one and, on the other hand, sufficiently simple to allow for the numerical
treatment. On top of that there is a problem of satisfying conditions (10).

This latter problem can be resolved by choosing the trial function as a
 function of arguments which are invariant under (10) by themselves. One 
of such arguments was proposed in [13]. Let $r=|z|$ and $r_1=|1-z|$. Then it 
is easy to see that
\beq
a=\frac{r^2r_1^2}{(1+r^2)(1+r_1^2)(r^2+r_1^2)}
\eeq
is invariant under (10). This is not the only  argument with this property.
Let us take
\beq
 b=\frac{(1-|z|^2)(1-|z_1|^2)(|z|^2-|z_1|^2)}{(1+r^2)(1+r_1^2)(r^2+r_1^2)}
\eeq
It is invariant under $z\rightarrow 1-z$ and changes sign
under
$z\rightarrow 1/z$, so that $b^2$ is invariant. Thus any function
$\Phi(a,b^2)$ will satisfy (10). To further narrow a possible choice of the
trial functions, one can impose
 condition that at $z\rightarrow 0$ it has the correct behaviour following
 from (14) and (15). Due to invariance under (10) it would mean that it would
 also have the correct behaviour near the other two singular points $z=1$ and
 $z=\infty$. For the ground state $h=1/2$ it means that at $z\rightarrow 0$
\beq
\Phi(z,z^*)\sim c_1r^{2/3}+c_2r^{5/3}(1+c_3\ln r+c_4\cos 2\phi)
\eeq
This suggests taking the trial function in the form
\beq
\Phi(a,b^2)= c_1a^{1/3}+c_2a^{5/6}(1+c_3\ln a+c_4b^2/a)
\eeq
In our calculations the trial functions were chosen in a more 
general form
\beq
\Phi=\sum_{k=1}^{N-N_1}c_ka^{k/2-1/6}+\sum_{k=1}^{N_1}d_ka^{k-1/6}\ln a+
fb^2/a^{-1/6}
\eeq
with $N+1$ variational parameters $c_k,d_k$ and $f$ (one of them is
 determined
by the normalization condition).
Note that the last term in $\Phi$ contains an  azimuthal dependence, which,
 due to the properties of $\epsilon_n(\nu)$ , can only raise the energy.
Therefore one cannot expect any improvement of the variational bound coming
 from it.
Indeed, our calculations incuding this term always lead to  $f$ close 
to zero for the minimizing function. Accordingly in the following we shall 
not discuss its influence and use (27) with $f=0$. 
 Function $\Phi$ with $N=3, N_1=1$ was 
used in the first variational
calculations in [13].

The basic quantity
$ \alpha $ given by Eq. (7) can  be
presented in the
form
\beq
\alpha_{n}(\nu)=(\frac{n}{2}-i\nu)
(\frac{n-1}{2}-i\nu)\big[(\frac{n+1}{2}+i\nu)f_{n}^{(1)}(\nu)+
(1-\frac{n}{2}-i\nu)f_{n-1}^{(0)}(\nu)\big]\eeq
where
\beq
f_{n}^{(k)}(\nu)=\int_{0}^{\infty}dr
r^{-1-k-2i\nu}\int_{0}^{2\pi}d\phi
e^{-in\phi}Z(r,\phi)
\eeq
Function
$
f_{n}^{(k)}(\nu) $ has  the properties:
$ f_{n}^{(1)}(\nu)=
f_{n}^{(0)}(-\nu)=(f_{n}^{(0)})^{*}$. Using them one can restrict  the summation
over $n$ and
 integration over
$ \nu $ to nonnegative values. 
 The
value of
the
$|\alpha_{n}(\nu)|^{2}
 $ can evidently be expressed via a single
function
$
f_{n}^{(0)}(\nu) $, which will simply be denoted  as
$ f_{n}(\nu) $ in the
following. In this manner one obtains
\beq
E=\sum_{n=0}^{\infty}\int_{0}^{\infty}d\nu\epsilon_{n}(\nu)
p_{n}(\nu)
\eeq
where for
$ n>0$
\[ p_{n}(\nu)=(\frac{n^{2}}{4}+\nu^2)(\frac{(n-1)^{2}}{4}+\nu^2)
\big[
(\frac{(n+1)^{2}}{4}+\nu^2)|f_{n}(\nu)|^{2}+
(\frac{(n-2)^{2}}{4}+\nu^2)|f_{n-1}(\nu)|^{2}+\]\beq
2\,{\mbox Re}\,
(\frac{n+1}{2}-i\nu)(\frac{2-n}{2}-i\nu)f_{n}(\nu)f_{n-1}(\nu)\big]
\eeq
and
\beq
p_{0}(\nu)=\nu^2(\frac{1}{4}+\nu^2)[(\frac{1}{4}+\nu^2)
|f_{0}(\nu)|^{2}+
(1+\nu^2)|f_{1}(\nu)|^{2}+
2\,{\mbox Re}\,(\frac{1}{2}-i\nu)
(1-i\nu)f_{0}(\nu)f_{1}(\nu)]
\eeq 
The
normalization functional
$ D $ has the same form (30) with
$\epsilon_{n}(\nu)\rightarrow 1 $.
Thus calculation of the
odderon energy
requires calculation of functions
$ f_{n}(\nu) $ and
$\epsilon_{n}(\nu)$.

The
main technical difficulty 
is the double Fourier transform (29). The
energy
$\epsilon_n(\nu)$ in $E$, Eq. (6), monotonously grows both with
$n$ and
$\nu$. It is negative only for $n=0$ and small enough values of
$\nu$. So the
problem with this formalism is that cutting in (5)
summation over $n$ and
integration over $\nu$ by some maximal values
$n_m$ and $\nu_m$, one always gets
 $E$ smaller than the exact value, corresponding to $n_m$ and
$\nu_m\rightarrow\infty$. Therefore in the
course of the calculation one always
approaches  the variational value
of $\epsilon$ {\it from below}.
 As we shall
see in the following, in fact,
rather high values of $n_m$ and $\nu_m$ are
necessary to obtain
$\epsilon$ with a good degree of accuracy.
This is the reason
why in [13]
a negative value  was obtained for $\epsilon$ (corresponding to
 $\alpha_O(0)\simeq 1.07$ ): too  small values of $n_m$ and $\nu_m$ were chosen
there. On the other hand, with
 high $n$ and
$\nu$, the double Fourier transform (29) becomes very difficult,
especially
having in mind that, due to the factors in (31), two first terms in
the
asymptotic expansion of $\alpha$ at high $n$ and $\nu$ cancel. As a result,
a trustworthy calculation
of $E$ and $D$ turns out to be very complicated, in
spite of its superficial transparency.

The important point in obtaining
reliable results has been using analytic asymptotic expansions for $f$ at high
$n$ and $\nu$, which are briefly discussed in Appendix.

\subsection{Numerical procedure}
As mentioned numerical integration in the double
Fourier transform (29) presents a formidable calculational task. 
As mentioned, our results were obtained by two groups working independently
 and using different choices of the number $N_1$ of logarithmic terms in (27)
 for
a given total number of terms $N$. We present here in some detail
the calculational procedure adopted in the computation with $N_1=1$.

In the integral (29) the integration over $r>1$ was transformed to $r<1$ by
a substitution
$ r\rightarrow 1/r $. The integration over $\phi$ was reduced to the interval
$0<\phi<\pi$ and $\exp(-n\phi)$ was substituted by $2\cos ( n\phi)$.
To soften the behaviour of the integrand at small values of $r$, three first
terms of its asymptotics at $r\rightarrow 0$ were subtracted and treated
in an exact manner. The final formula for $f_n(\nu)$ used in the calculations
 was thus
 \[f_{n}(\nu)=2\int_{0}^{\pi}d\phi
\cos in\phi\int_{0}^{1}dr
\big(r^{-2i\nu} 
(\zeta(r,\phi)-\zeta_{0})\]\[
+r^{-1+2i\nu}(\zeta(r,\phi)-\zeta_{0}+\zeta_{1}r
\ln
r-\zeta_{2}r)\big)+\]
\beq\delta_{n0}2\pi
(\frac{\zeta_{0}}{2i\nu(1-2i\nu)}+
\frac{\zeta_{1}}{(1+2i\nu)^{2}}+\frac{\zeta_{2}}{1+2i\nu}
\eeq
where
$\zeta=Z/r$ and the subtraction constants are
\beq
\zeta_{0}=c_1(1/2)^{1/3},\ \zeta_{1}=-2c_{N}(1/2)^{5/6},\
\zeta_{2}=
(1/2)^{5/6}(c_{2}-c_N\ln 2)
\eeq
Eq. (33) was used for numerical calculation of
$ f_{n}(\nu) $ in the  
interval of $0\leq n<30$ and $0<\nu<15$.
Integrations were performed by dividing
the rectangle
$0<r<1,\ 0<\phi<\pi$ into an $M\times M$ grid, interpolating
$\zeta$
quadratically on the grid and then doing the integrals explicitly. The
maximal value of $M$ was 640. The achieved accuracy was about $10^{-5}$.
 Thus
calculated values of $f_n(\nu)$ were summed over $n$ and integrated over $\nu$
as indicated in (30)-(32) to obtain $E$ and $D$.
Stable results were obtained
with the quite high maximal values $n_m=300$ and $\nu_m=150$.
In the part of
($n,\nu$) space outside the rectangle
$0\leq n<30,\ 0<\nu<15$
 the asymptotic
expressions were used for $f_n(\nu)$ (see Appendix). 
As a result we calculated
$E$ and $D$ as a quadratic form in the variational
parameters $c_k,d_k$. Afterwards
the minimal value $\epsilon$ of $E$, subject to condition $D=1$, was found by
standard methods. 

Other calculations used (27)  with $N_1=1$ for $N=3,4$ and $N_1=2$
for $N=6$.
The adopted numerical procedure was
different, however the  final results, as we shall presently see, are quite
similar.  

Our results for  different number of parameters $N$ are presented in 
Table 1 for the both choices: $N_1=1$ always and $N_1=1$ for $N=3,4$,
$N_1=2$ for $N=6$. The
corresponding energies are denoted $\epsilon_1$ and $\epsilon_2$ respectively.
Adjacent columns present values of the variational parameters for both cases
($c_1=1$).
 The standard precision corresponds
to the ($r,\phi$) grid 320$\times$320.
 To clarify the accuracy achieved we also
present the results for $\epsilon_1$ with
a double precision (the grid 640$\times$640) for
$N=5$.

Inspecting these
results we see that the final accuracy in energy
is of the order
$5.10^{-3}$. Also taking $N>6$ with $N_1=1$ leads to no improvement
within the precision achieved, since the corresponding change in energy is  of
the
 same
order or less. So our conclusion is that the variational odderon energy with a
trial function (27) is given by (3) and  that with the accuracy  achieved in
the
course of numerical integration, as described above, the maximal number
of
terms to be taken in the trial function is $N=6$, although already with
$N=3$ 
the energy is obtained up to 1\%.

\section{Exact wave functions}
\subsection{Problems of precision}
In this section we report on the calculations of the odderon energy
using  Eqs. (4) and (5) with the odderon wave function (15)
 obtained after solving Eq. (21).
These functions can be obtained in the form of power series in $z$ and
$z^*$, different in different parts of the complex $z,z^*$ plane. The exact
solution is, of course, analytic in each of the variables, except at the
 three mentioned singular points $z=0,1,\infty$. So it is absolutely smooth
 at the
boundaries of the regions in which it is represented by different
 power series. However, in practice one knows it to some finite precision.
 This causes
certain discontinuities in the function itself and its derivatives at the
mentioned boundaries. At high values of $n$ and $\nu$ the double Fourier
 transform (29) is very sensitive
to such local irregularities of the wave function. They result in a very
 poor precision for the calculated odderon energy and even in a completely
 wrong
order of magnitude for it. Because of this,    
 our first step was to
obtain the odderon wave function with a higher precision
 than
reported in [9].
To this end we set up a program which essentially repeats
the procedure
employed in [9] and allows to reduce discontinuities of
$\Phi$ calculated in different variables to values of the order $10^{-9}$.
The
only difference  is that we used the basic functions $u_i$,
$i=1,2,3$,
multiplied by appropriate factors to make them real at points where the transfer
matrices are calculated. This substantially
facilitates achieving the desired
accuracy.

 For the eigenvalue $q_3$ and wave
function parameters $\alpha$,
$\beta$ and $\gamma$ we obtained the following
values (precision
$10^{-9}$)
\beq iq_3=0.205257506,\ \ 
\alpha=0.709605410,\ \
\beta=-0.689380668,\ \ 
\gamma=0.145651837\eeq
However even with these high
precision values direct calculation of $\Phi$
in one of the three sets of
variables $z$, $1-1/z$ or $1/(1-z)$ fails in the
vicinity of the point
 $z_0=\exp
(i\pi/3)$ where none of the series
converges absolutely. In spite of the fact
that $z_0$ is only a single point in the $(r,\phi)$ plane, this makes it
practically
impossible  to calculate the double Fourier transform for $|n|>10$
and/or $|\nu |>5$.
To overcome this difficulty we had to redevelop the
function
$\Phi$ around the intermediate point $(1/2)z_0$. With this
redevelopment
100 terms in the series proved to be sufficient to obtain reliable
results.

The Fourier transform  was again performed by interpolating
$Z$
quadratically on the $(r,\phi)$ grid and doing the integrals in $r$ and
$\phi$ analytically. Reasonable results are obtained already with a 160 x
160
grid. We however also used  320 x 320 and 640 x 640 grids to
analyse the
precision achieved at this step.

Even with a 640 x 640 grid the numerical
Fourier transform becomes
unreliable for $|n|>30$ and/or $|\nu |>15$. For such
high values of
$|n|$ and $|\nu|$ we used asymptotic formulas for the Fourier
transform,
which can easily be obtained from the expansion of $\Phi$ around
$z=1$ in a
 similar manner as for variational wave functions in the preceding section
 (see Appendix).

Our final cutoffs were chosen to be $|n|<300$ and $|\nu|<150$, which proved to
be quite sufficient for the determination of $\epsilon$
with a precision 
$0.001$.

\subsection{Ground state energy} Results of our numerical calculation
of $D$
 and $E$ in
the region $|n|<30$ and $|\nu|<15$ using an $N$ x $N$ grid in the
$(r,\phi)$ plane are presented in the Table 2 for $N=160,320$ and
$640$.

To
these values one has to add the contributions from the asymptotic
region
described in the preceding section. They are

\beq
\Delta D=0.002398,\ \ \Delta
E=0.018451
\eeq
Taking the results at $N=640$ as the most accurate ones we
finally have
\beq
D=1.644454,\ \ E=0.273083,\ \ \epsilon=0.1660
\eeq Thus our
result for $\epsilon$ coincides with the value found in [9]
up to $10^{-3}$.
With all the difficulties involved in the numerical
calculations, we  consider
this agreement  quite satisfactory.
So direct calculation of the odderon energy
confirms the result found
by JW. 

\subsection{Higher eigenvalues of $q_3$}
 We have  tried to check the
result of [9] also for the excited
state with the next higher value of $iq_3$.
Unfortunately in this
case calculations proved to be still more difficult and we
could not arrive at  a result of a convincing accuracy.

Our precise
calculations  of the wave function gave for this state:
\beq
iq_3=2.343921063,\
\ \alpha=0.391855163,\ \ 
\beta=-0.0533712012,\ \
\gamma=0.918477570
\eeq
Numerical calculation of $D$ and $E$ in the
region
$|n|<15$, $|\nu|<15$ gave results presented in Table
3.
As one observes, the achieved accuracy does not
exceed 15\%. Analyzing
these numbers one can see that all the error comes from
the region of
maximal $|n|$ and $|\nu|$ where the double Fourier transform
is
apparently performed inaccurately.
>From these numbers we can only conclude
that for this excited state
\beq
\epsilon\simeq 2.0\pm 0.3
\eeq
The value found
in [9] is 1.7231... Our result does not contradict
this number.
\subsection{The case $q_3=0$} We have also studied a degenerate solution for
 $q_3=0$, discussed in [9].
 However in this case we were not able to construct a
wave function unique in the $(r,\phi)$ plane and satisfying the
necessary boundary conditions and symmetry requirements. At $q_3=0$ the solution
$u^{(0)}_3(z)$ has no logarithmic term.
 If, following [9], we seek $\Phi$ in the
form (15) and require the same
conditions (16) to be satisfied, then Eq. (21) has no solutions at all, which
means that the Bose symmetry cannot be fulfilled. The problem is related to the
fact, that, with the logarithmic term missing in $u_3$, a unique solution is
obtained also with
$A_{33}\neq 0$. As a result, starting from a solution with $A_{33}=0$
one always obtains $A_{33}\neq 0$ after applying the transformation (21).

One may wonder if a solution containing a product $\bar{u}_3(z^*)u_3(z)$
is admissible as a physical odderon wave function. Such a solution 
does not vanish as the distance between any of the three gluons becomes small.
 This can be easily
seen from the behaviour at $z\rightarrow 0$ of the complete  wave function
(9): the explicit factor $r^{2h/3}$ cancels against the leading term
 $r^{-2h/3}$
in the product  $\bar{u}_3(z^*)u_3(z)$ and one gets a constant.
In fact the solution with this property can be easily constructed explicitly
[12]. It is just a sum of three pomeron wave functions
\beq
\Psi(r_1,r_2,r_3)=\Psi_{BFKL}(r_1,r_2)+\Psi_{BFKL}(r_2,r_3)+
\Psi_{BFKL}(r_3,r_1)
\eeq
This function is an evident eigenfunction of $\hat{q}_3$ for zero eigenvalue
and it is conformal and Bose symmetric. However it cannot be considered as a
physical odderon state, since the Hamiltonian $H$ cannot be applied to it due
 to its singularities at low gluon momenta. Take the first term, which does
 not depend on $r_3$. Then it is proportional to $\delta^2(k_3)$ in the
 momentum space $k_1,k_2,k_3$ of the three gluons. The part of the odderon
 Hamiltonian depending on $k_3$ contains a term proportional to $\ln k_3^2$
 and two interactions
between the gluons 31 and 32. It can be easily seen that the latter give
finite result, applied to a wave function containing $\delta^2(k_3)$.
The $\ln k_3^2$, however, is infinite. So $H$ applied to (40) is not defined.
This property can also be seen from the approximate relation
 $H\simeq\ln\hat{q}_3$ valid for small values of $z$ [5]. Evidently $H$
 diverges as $\hat{q}_3\rightarrow 0$.

Thus our conclusion is that the $q_3=0$ state
does not
correspond to any physical odderon state.\vspace{0.8cm}

\subsection{"Moving" odderon}
For conformal weights $h=\frac{1}{2}+i\sigma$
 the
odderon energy
is supposed to behave at small $\sigma$
as
\beq
\epsilon(\sigma)=\epsilon_0+a\sigma^2
\eeq where $\epsilon_0$ is the
value (2) and $a$ is a parameter which
determines the diffusion of the odderon
wave function in the momentum space. This parameter has long been  known for the
pomeron to be $14\zeta(3)\sim 16.8$
(in units $3\alpha_s/\pi$). It is of certain
interest to find $a$ for the
 odderon.
To this aim we first found the parameters of the odderon wave
function
for various (small) $\sigma$ using the same method as employed
for
$h=1/2$. Our results are presented in Table 4. The value of $iq_3$
turned
out to be real for arbitrary $\sigma$, whereas, with $\alpha$ chosen
to
be real, both $\beta$ and $\gamma$ result complex. We chose
$\alpha=1$.

 Inspecting these figures one immediately notes that
$|\gamma|=iq_3$. This relation was predicted  by L.N.Lipatov
[15].

With the odderon parameters found, we
calculated the odderon  energies
directly,
using the same techique as for $\sigma=0$. With $\sigma$
different
from zero calculation becomes still more cumbersome and time
and
memory consuming due to lack of certain symmetries and overall
complex arithmetics. For these reasons we had to limit ourselves with a maximal
160 x 160 grid in the $(r,\phi)$ plane and neglected the
contribution from the
asymptotical region $n>30$, $|\nu|>15$.
Our results are shown in Table 5
together with the ones  obtained
via the solution of the Baxter equation
[16].

Our
energies lie a little below the ones obtained from the Baxter equation,
which is
natural since we have neglected the
asymptotic part  of the $n,\nu$. Having this
in mind we find a complete agreement between our direct calculation results
and
the ones based on the Baxter equation.
>From our energies  we find for the
parameter $a$ in (41)
\[  a=0.61.
\]
More precise energies found in [14] lead
to
\[
a=0.605.\]
Note however that already at $\sigma=1$ the approximation
(41)
breaks down and more powers of $\sigma^2$ are needed to describe the
energy
behaviour.
It is interesting that the parameter $a$ for the odderon is much
smaller than
the one for the pomeron. In fact their ratio is of the same order
as the
ratio of corresponding energies.

\section{Conclusions}
Our calculations confirm the results obtained by JW for the perturbative
odderon intercept and thus seem to remove any doubts concerning the validity
 of their procedure to
relate the $\hat{q}_3$ operator and the Hamiltonian.

The variational calculations give a result for $\epsilon$
which is 
$\sim$30\% larger than the exact value. However they also convey the
important message that the intercept of the odderon 
lies quite close to unity being slightly smaller.
In fact, for realistic values of $\alpha_s$, the intercept $\alpha_O(0)$
 calculated variationally is at most 2\% lower than the exact one: 0.94
 instead of 0.96.

The disputed eigenvalue $q_3=0$ does not seem to correspond to any physical
odderon state.

In the course of the variational calculations a simple approximate form
of the odderon wave function is found, which allows a realistic
calculation of the odderon residues,  important for the study of processes
involving the odderon exchange.

\section{Acknowledgements}
The authors express their  gratitude to  J.Wosiek,
whose
comments suggested  and accompanied part of this investigation
and helped in improving the manuscript,
 and to  L.N.Lipatov
and G. Korchemsky
for very interesting and illuminating discussions. They are  grateful to all
of
them for communicating their yet unpublished results.
M.A.B. is thankful to
LPTPE, Universit\'e  Pierre et Marie Curie, for  hospitality during his stay in
Paris, where this paper has been completed.

\section{Appendix. Asymptotics at large $n$ and $\nu$ for variational
 calculations}
Passing to the variable
$\rho=-\ln r$ and introducing
2-dimensional
 vectors $x=(\rho,\phi)$
and $w=(z,n)=(2\nu,n)$ we rewrite (29) as
\beq
f_n(\nu)=\int_{-\infty}^{+\infty}d\rho\int_{-\pi}^{+\pi}d\phi
e^{iwx}Z(x)\eeq
The integration
point $x=0$ is obviously essential  for the asymptotics at high
$n$ and
 $\nu$.
At $x\rightarrow 0$, keeping terms up to third order in small $\rho$
 and
$\phi$, we have
($x=\sqrt{\rho^2+\phi^2}$):
\beq Z_p=(rr_1)^{1/3}a^p=(1/2)^p x^{2p+1/3}(1-(1/2)\rho
+a_1\rho^2+b_1\phi^2+
c_1\rho^3+d_1\rho\phi^2)\eeq where
\[a_1=5/36-(29/12)p\ \ ,\ \ b_1=-1/72-(25/12)p\ \ ,\]
\[
 c_1=-1/36+(29/24)p\ \ ,\ \
d_1=1/144+(25/24)p\] and  \[p=k/2-1/6\ \ , \ \  k=1,2,\ldots\] 
For the term with a logarithm, in 
the same manner we obtain
\[\tilde{Z}_p=(rr_1)^{1/3}a^p\ln a=(1/2)^p
x^{2p+1/3}[\ln
(x^2)(1-(1/2)\rho
+a_1\rho^2+b_1\phi^2+c_1\rho^3+d_1\rho\phi^2)+\]\beq
\ln(1/2)-(1/2)\ln(1/2)\,\rho+
a_2\rho^2+b_2\phi^2+c_2\rho^3+
d_2\rho\phi^2)]\eeq
where
\[p=k-1/6\ \ ,\ \ k=1,2,\ldots \]
and
\[a_2=a_1\ln(1/2)-29/12\ \ ,\ \ b_2=b_1\ln(1/2)-25/12\ \ ,\]
\[c_2=c_1\ln(1/2)+29/24\ \ ,\ \ d_2=d_1\ln(1/2)+25/24\]
Inserting these expressions into the integral (42) and
extending the
integration over $\phi$ to the whole real axis one obtains
the
asymptotical expansion of different terms in $f_n(\nu)$. In particular the
asymptotical expansion of the term originating from
$Z_p$ is found
as
\[f_n^{(p)}(\nu)=c_p(1+(1/2)id/dz-ad^2/dz^2-bd^2/dn^2+icd^3/dz^3+id
d^3/dzdn^2)w^{-\alpha-1}\]
where
$\alpha=2p+4/3$ and
\[c_p=2^{\alpha+1-p}\Gamma^2(1/2+\alpha/2)\cos
(\pi\alpha/2)\]
Doing
the derivatives, one obtains finally ($w=\sqrt{n^2+4\nu^2}$):
\beq {\rm
Re}
f_n^{(p)}(\nu)=c_pw^{-\alpha-1}\left(1+(\alpha+1)w^{-4}[z^2(b-a(\alpha+2)
+n^2(a-b(\alpha+2))]
\right)\eeq
\[ {\rm
Im} f_n^{(p)}(\nu)=-(1/2)c_p(\alpha+1)zw^{-\alpha-3}\]\beq
\left(1
+2(\alpha+3)w^{-4}[z^2(c(\alpha+2)-d)+n^2(d(\alpha+4)-3c)]\right)\eeq

For the term with a logarithm  only the part with $\ln
x^2$
contributes. The result coincides with the formula above where
the
constant $c_p$ is substituted by
$16\pi (1/2)^{p}$.If one puts these
asymptotic expressions into (31) one finds that the two leading terms coming
from $Z_{1/3}$ cancel. Numerically the asymptotic
expansion begins to work at
rather high values of $n$ and
$\nu$:
$\sqrt{n^2+4\nu^2}>30$.

\section{References}

\noi 1. H1 Collab., S.Tapprogge et al., in Proc. of the Int. Conf. on
the Structure and the Interactions of the Photon (Photon 92), World. Sci.,
Ed. F.C.Erne and A.Bujis.

\noi 2. I.F.Ginzburg and D.Yu.Ivanov, Nucl. Phys. {\bf B388} (1992) 376;\\
  J.Czyzewski, J.Kwiecinski, L.Motyka and M.Sadzikowski,
Phys. Lett. {\bf B398} (1997) 400; {\bf B411} (1997) 402;\\
  R.Engel, D.Ivanov, R.Kirschner and L.Szymanowski, Eur. Phys. J. {\bf C4}
(1998) 93;\\
 L.Motyka and J.Kwiecinski, hep-ph/9802278.

 \noi 3.  W.Kilian and O.Nachtmann, hep-ph/9712371, to be published in
 Eur. Phys. J. {\bf C};\\
 M.Rueter, H.G.Dosch and O.Nachtmann, hep-ph/9806342.

\noi 4. J.Bartels, Nucl. Phys. {\bf B175} (1980) 365;\\
 J. Kwiecinski and M.Praszalowicz, Phys. Lett. {\bf B94} (1980)
413.

\noi 5. L.N.Lipatov, Sov. Phys. JETP {\bf 63} (1986) 904;
 Phys. Lett. {\bf B251} (1990) 284.

\noi 6. L.N.Lipatov, Phys. Lett. {\bf B309} (1993) 394.

\noi 7. L.N.Lipatov, JETP Lett, {\bf 59} (1994) 571;\\
L.D.Faddeev and G.P.Korchemsky, Phys. Lett {\bf B342} (1994) 311;\\
G.P.Korchemsky, Nucl. Phys. {\bf B443} (1995) 255; {\bf B462} (1996) 333.

\noi 8. J.Wosiek and R.A.Janik, Phys. Rev. Lett. {\bf 79} (1997) 2935.

\noi 9. R.A.Janik and J.Wosiek, Krakow preprint TPJU-2/98, 
hep-th/9802100; see also L.N.Lipatov, talk given at IPN Orsay,France,March
1997.

\noi 10. M.Praszalowicz and A.Rostworowski, hep-ph/9805245.

\noi 11. N.Armesto and M.A.Braun, Santiago preprint US-FT/9-94,
hep-ph/9410411; Z.Phys. {\bf C63} (1997) 709.

\noi 12. G.P.Korchemsky, {\it private communication.}

\noi 13. P.Gauron, L.N.Lipatov and B.Nicolescu, Phys. Lett. {\bf B304}
(1993) 334; Z.Phys. {\bf C63} (1994) 253.

\noi 14. P.Gauron, L.N.Lipatov and B.Nicolescu, Phys. Lett. {\bf B260}
(1991) 407.

\noi 15. L.N.Lipatov, hep-ph/9807477.

\noi 16. J.Wosiek, {\it private communication.}

\newpage

\begin{center}{\bf Table 1. Odderon energy and parameters of the trial

functions}\vspace{1.cm}
\begin{tabular}{|r|c|l|c|l|}\hline
N&$\epsilon_1\
(N_1=1)$&$ c_2,c_3,...;
d_1$&$\epsilon_2(N_1=1,2)$&$c_2,c_3,...;d_1,d_2$\\\hline
3&0.22865&-0.5036;
0.2895& 0.23137& -0.50276;0.28936 \\\hline
4&0.22632&-0.2791,-0.3190;0.3609&
0.22735&-0.21420,-0.41031;0.38136\\\hline
5&0.22627&-0.2009,-0.5052,0.1557;& &  
\\ &       & 0.3779       &   &            
\\\hline
$5^*$&0.22634&-0.2021,-0.5028,0.1543;& &       \\  &       
&0.3775      &   &                 \\\hline
6& 0.22619&-0.3735,0.08842,-0.9765,
&
0.22269&0.49231,-3.49272,-1.82821;\\ &        &1.003;0.3471   & 
&0.50316,-1.49528                 \\\hline 7& 0.22618&    &  
&                    \\\hline8& 0.22616&    &   &                   
\\\hline
9& 0.22616&    &   &                    
\\\hline
\end{tabular}
\end{center}
\hspace{3.cm}$^*)$ double precision.\vspace{1 cm}
  
\begin{center}
{\bf Table 2. $D$ and $E$ for the ground
state}\vspace{0.5cm}

\begin{tabular}{|r|c|c|}\hline
N& D&
E\\\hline
160&1.642162&0.255480\\\hline 320&1.642085&0.254852\\\hline
640&1.642056&0.254632\\\hline
\end{tabular}
\end{center}

\newpage
\begin{center}
{\bf Table 3. $D$ and $E$ for the state with
$iq_3=2.343921063$}
\vspace{0.5cm}

\begin{tabular}{|r|c|c|}\hline
N& D&
E\\\hline
160&2.92863&6.08989\\\hline   
320&2.80693&5.27381\\\hline
640&3.05939&6.96764\\\hline
\end{tabular}
\end{center}
\vspace{1cm}

\begin{center}
{\bf Table 4. Odderon parameters for
$h=\frac{1}{2}+i\sigma$}\vspace{0.5cm}

\begin{tabular}{|r|c|c|c|}\hline$\sigma$&$iq_3$&$\beta$&$\gamma$
\\\hline   
0.01&0.205306079&-0.971740164-i0.014404102&0.205305637-i0.000425478 
\\\hline
0.1&0.210089247&-0.995153863-i0.142974530&0.210052319-i0.003938872    
\\\hline
0.3&0.247227544&-1.156524786-i0.415163678&0.247186043-i0.004529717 
\\\hline 
0.5&0.316528176&-1.395571390-i0.695891904&0.316214188+i0.014095104   
\\\hline
1.0&0.619239545&-2.044631201-i1.672240784&0.591391973+i0.183611401
\\\hline
\end{tabular}\end{center}
\vspace{1cm}

\begin{center}
{\bf Table 5. Odderon energies for $h=\frac{1}{2}+i\sigma$
}\vspace{0.5cm}

\begin{tabular}{|r|c|c|}\hline
$\sigma$&$\epsilon$ & $\epsilon$
[14] \\\hline
0.0&0.1534&0.16478\\\hline 0.1&0.1597&    
\\\hline
0.3&0.2085&0.21777\\\hline
0.5&0.2980&0.30523\\\hline
1.0&0.6269&0.63228\\\hline
\end{tabular}
\end{center}
\end{document}